# Determination of band alignment in the single layer MoS$_2$/WSe$_2$ heterojunction

Ming-Hui Chiu[1,^], Chendong Zhang[2,^], Hung-Wei Shiu[3], Chih-Piao Chuu[1], Chang-Hsiao Chen[1], Chih-Yuan S. Chang[1], Chia-Hao Chen[3], Mei-Yin Chou[1,4,5], Chih-Kang Shih[2,*] and Lain-Jong Li[1,6,*]

[1]Institute of Atomic and Molecular Sciences, Academia Sinica, No. 1, Roosevelt Rd., Sec. 4, Taipei 10617, Taiwan
[2]Department of Physics, University of Texas at Austin, Austin, TX 78712, USA
[3]National Synchrotron Radiation Research Center, HsinChu 30076, Taiwan
[4]School of Physics, Georgia Institute of Technology, Atlanta, GA 30332, USA
[5]Deapartment of Physics, National Taiwan University, Taipei 10617, Taiwan
[6]Physical Sciences and Engineering Division, King Abdullah University of Science and Technology, Thuwal, 23955-6900, Kingdom of Saudi Arabia.

^These authors contributed equally.
*Corresponding author E-mail: lance.li@kaust.edu.sa, shih@physics.utexas.edu
.

**The emergence of transition metal dichalcogenides (TMDs) as 2D electronic materials has stimulated proposals of novel electronic and photonic devices based on TMD heterostructures. Here we report the determination of band offsets in TMD heterostructures by using microbeam X-ray photoelectron spectroscopy (μ-XPS) and scanning tunneling microscopy/spectroscopy (STM/S). We determine a type-II alignment between MoS$_2$ and WSe$_2$ with a valence band offset (VBO) value of 0.83 eV and a conduction band offset (CBO) of 0.76 eV. First-principles calculations show that in this heterostructure with dissimilar chalcogen atoms, the electronic structures of WSe$_2$ and MoS$_2$ are well retained in their respective layers due to a weak interlayer coupling. Moreover, a VBO of 0.94 eV is obtained from density functional theory (DFT), consistent with the experimental determination.**

Transition metal dichalcogenides (TMDs) have emerged as a new platform for atomic layer electronics[1] and optoelectronics[2-5]. Many proposed novel devices are based on heterostructures formed between dissimilar TMDs[6-9]. Heterojunction band offset (HJBO) is the key parameter for designing HJ-based electronic/photonic devices and accurate determination of this parameter is of critically important. In conventional semiconductor HJs, one commonly used technique to determine the valence band offset (VBO) is XPS[10-13]. This method relies on finding the core-level alignment of two constituent semiconductors across the heterojunction (HJ). Then with additional information on the core-level position relative to the valence band maximum (VBM) measured separately for individual semiconductors, the VBM alignment across the HJ can be determined. The application of this technique to TMD HJs, however, faces two technical challenges. First, the HJ is formed only locally with a lateral length scale of only a few microns, owing to the limited lateral size of available TMD monolayer samples. It is thus necessary to locate such locally formed HJs and measure the core level alignment across the junction. We overcome this challenge by using µ-XPS where the photon spot can be focused down to sub-microns (spot size ~ 100 nm) to measure all the relevant quantities at the local scale. The second challenge is the determination of the VBM position. In conventional semiconductors the valence bands primarily comprise of *sp* orbitals, with a smooth density of states (DOS). This allows for a precise determination of the VBM through curve fitting. In single layer (SL) TMDs, the density of states (DOS) near VBM has a complicated line shape due to the different characteristics of the states near Γ and K. We overcome this difficulty by using STM/S measurements of individual TMDs to determine the quasi-particle gaps and the fine structures involved, such as the energy difference between the VBM at Γ and K. The special combination of µ-XPS and STM/S measurements allow us to deduce the precise band offset values including both VBO and CBO.

In addition, the first-principles calculations are performed to elucidate the interlayer interaction.

**Results**

For μ-XPS measurements, the single-crystalline monolayer TMDs were synthesized using chemical vapor deposition (CVD) on sapphire substrates[14-16]. These as-grown TMDs were then detached from sapphire substrates and transferred onto Si wafers with native oxides to form the HJ stacks (see the Methods section). It has been shown recently that with the presence of a thin native oxide (~ 2 nm) layer the photoemission measurement can be carried out without the charging effect, but the oxide is also thick enough to ensure that the Si band structure is suppressed so the valence band structure detected is only from SL-TMDs[17]. For STM measurements, SL-TMDs are grown on highly-oriented-pyrolytic-graphite (HOPG) directly to provide enough conductivity for STM imaging.

Figure 1a shows the optical micrograph and atomic force microscopy (AFM) images for the $WSe_2/MoS_2$ heterostructure stacked on a sapphire substrate. The characterizations by AFM, optical gap measurements, and Raman spectroscopy (Supplementary Fig. S1 and Table S1) indicate that they are single-layer TMDs flakes[18,19]. Fig. 1b shows the photoluminescence (PL) spectra for the selected sites including $MoS_2$ only (A), $WSe_2$ only (C) and $WSe_2/MoS_2$ (B) areas. The PL intensity of both $MoS_2$ and $WSe_2$ for the overlapped area (B) is significantly lower than that from $WSe_2$ or $MoS_2$ alone, indicating that the photoexcited carriers are quenched through other routes than the emission from individual $WSe_2$ or $MoS_2$ band edges. The surface adsorbates on these 2D materials are removed by vacuum annealing at an elevated temperature[19]. Herein, these flakes are annealed in a high vacuum chamber ($2 \times 10^{-10}$ Torr) at 300 °C for over 8 hours prior to the scanning photoelectron microscopy (SPEM) scans.

Figure 2a shows the spatial mapping of Mo3d$_{5/2}$ (left), W4f$_{7/2}$ (middle) and their composite (right) respectively using µ-XPS. Such an elemental mapping allows us to identify isolated SL-WSe$_2$, SL-MoS$_2$, and locally stacked HJ bilayer unambiguously. The corresponding core level spectra of W4f and Mo3d in isolated SL-WSe$_2$, isolated SL-MoS$_2$, the WSe$_2$/MoS$_2$ stack and the MoS$_2$/WSe$_2$ stack are shown in Fig. 2b. Also shown are the corresponding valence band (VB) spectra in isolated SL-WSe$_2$ and SL-MoS$_2$ flakes. The energy splittings and relative intensities due to the spin-orbit coupling for W4f (4f$_{7/2}$ and 4f$_{5/2}$) and Mo3d (3d$_{5/2}$ and 3d$_{3/2}$) doublets remain unchanged before and after forming 2D stacked films. Furthermore, their energy locations are independent of the X-ray beam flux, confirming the absence of the charging effect.

The core level separation between the W4f$_{7/2}$ and Mo3d$_{5/2}$ is determined to be 196.97 ± 0.04 eV across the HJ. In reverse stacking (*i.e.* WSe$_2$/MoS$_2$) this value becomes 196.98 ± 0.04 eV, essentially unchanged within the experimental error. In order to determine the VBO, we also need the respective quantities of core-level relative to the VBM in individual SL-WSe$_2$ and SL-MoS$_2$. With a least square fit of the leading edge of the VB spectra (shown in Fig. 2b), the apparent VBM locations are determined to be 0.92 eV for WSe$_2$ and 1.11 eV for MoS$_2$ (Supplementary Fig. S2). We recognize, however, that this fitting procedure may not lead to the correct value of VBM value due to the complicate line shape in the DOS (discussed further in Supplementary Information). We thus label this "apparent" VBM position as VBM*. Using this procedure, we determine that the Mo3d$_{5/2}$ relative to the VBM* in MoS$_2$ to be 228.33 ± 0.04 eV and the W4f$_{7/2}$ relative to the VBM* in WSe$_2$ to be 31.77 ± 0.04 eV. Assuming that the energy difference between the VBM* and the core level remains unchanged after stacking, the apparent valence band offset (VBO*) for MoS$_2$ and WSe$_2$ and in the stacked film MoS$_2$ on WSe$_2$ can be determined as:

$$\Delta E_{v*}^{MoS_2-WSe_2} = 31.77 \text{ eV} + 196.97 \text{ eV} - 228.33 \text{ eV} = 0.41 \text{ eV}$$

Here the notation *MoS₂ – WSe₂* in the superscript indicates that this is the potential step moving from MoS$_2$ into WSe$_2$. The error bar in the determination of the VBO* is estimated to be 0.07 eV. The consistent core level separation of 196.98 ± 0.04 eV between W4f$_{7/2}$ and Mo3d$_{5/2}$ in reverse stacking also implies the validity of VBO commutativity in TMD HJs.

Theoretical calculations show that the DOS near the VBM is dominated by the states near the Γ point with much less contribution from those near the K point where the true VBM is located (as shown in supplementary Fig. S3b, and the generic band structures can also be seen in the Fig S3a.). Since XPS is unable to resolve the Γ-K splitting, the measured VBM* likely corresponds to the location of the Γ point. Thus, the resulting VBO* would not correspond to the true VBO. This short-coming is overcome by using STS to accurately determine the quasiparticle gaps, and the Γ-K VBM energy splitting, $\Delta_{\Gamma-K}$. With these additional pieces of information, the band alignment (both valence band and conduction band) can be accurately determined.

Figure 3a shows the STM image of SL-WSe$_2$ grown on HOPG, with a typical tunneling spectrum shown in Fig. 3b. In the negative bias range there is sharp peak at about -1.65 V then vanishes quickly as the bias is increased. In the positive sample bias range, an onset of conductivity occurs at +1.03 V. In order to determine the VBM and CBM positions, the conductivity is displayed in the logarithmic scale (lower panel in Fig. 3b) where the sharp onset at 1.03 V can be unambiguously identified as the CBM. The determination of the VBM using such a *dI/dV* spectrum, however, encounters some ambiguity. An extrapolation of the leading edge would yield a threshold being located somewhere between -1.4 and -1.3 V. Detailed examination, however, reveals that the global VBM at the K point is located at a significantly higher energy location (at -1.05 V) and the peak at the -1.65 V corresponds to the local VBM at the Γ point. In the WSe$_2$ region, as long as the junction stabilization voltage (namely, the applied bias before the interruption of the feedback) is between - 1.65 and - 2.0 V, the spectrum does not have enough sensitivity to detect the global VBM position. To enhance the ability to observe these states, we acquire another spectrum by reducing the tip-to-sample distance by ~ 3.5 Å, shown as the red curve in Fig. 3c. In this "close-in" distance, tunneling from the states of the underlying graphite can also be detected. Moreover, there is a threshold at -1.05 ± 0.05 V separating the states derived from graphite and those from WSe$_2$. This threshold is assigned as the global VBM at the K point. By using a stabilization voltage of - 0.8 V (well into the gap region of WSe$_2$), the underlying graphite electronic states can be "seen through SL-WSe$_2$" and the spectrum is essentially the same as the spectrum acquired in the bare graphite surface (shown as the black and blue curves in Fig. 3c). This also confirms our assignment of the threshold at -1.05 V to be the global VBM position at K point. More detailed discussions on the determination of energy location of different thresholds and their origins in the Brillouin Zone (BZ) can be found in ref. 20. In Fig. 3d we show a close-in spectrum for MoS$_2$ where the Γ point can be identified at -2.0 V while the VBM is located at -1.84 ± 0.05 eV. The CBM location of MoS$_2$, can also be resolved as 0.31 ± 0.05 eV.

In order to determine the core-level position relative to the true VBM in SL-TMD, we carry out µ-XPS on similarly prepared single layer MoS$_2$ and WSe$_2$ on HOPG (Supplementary Fig. S4). The measurements yield a binding energy of 229.24 ± 0.03 eV for Mo3d$_{5/2}$, corresponding to a separation of 227.40 eV to the true VBM (at the K point) in MoS$_2$. Similarly W4f$_{7/2}$ in SL-WSe$_2$ on graphite has a binding energy of 32.31 ± 0.03 eV, corresponding to a value of 31.26 eV when referenced to the true VBM. With these two binding energies relative to their individual VBM determined, we can re-apply the same algorithm and obtain a true VBO of

$$\Delta E_v^{MoS_2-WSe_2} = 31.26 \text{eV} + 196.97 \text{ eV} - 227.40 \text{ eV} = 0.83 \pm 0.07 \text{ eV}.$$

One can immediately see that the VBO measured is 0.83 eV larger than the apparent VBO* of 0.41 eV measured using XPS alone. This is due to the fact that in WSe$_2$, the K point is 0.60 eV higher than the Γ point while in MoS$_2$, it is only about 0.16 eV higher. With these STS energy data, the VBO* value can be deduced as 0.39 eV, very close to the XPS values of 0.41 eV. This affirms our earlier conjecture that the XPS measurement of VBM* is dominated by the high DOS at the Γ point. STS measurements of quasiparticle gaps (2.15 ± 0.1 eV for MoS$_2$ and 2.08 ± 0.1 eV for WSe$_2$) also allow us to deduce a conduction band offset (CBO) of 0.76 ± 0.12 eV, affirming a type II band alignment as illustrated in Fig. 3e.

Another very interesting observation is that for individual SL MoS$_2$ and WSe$_2$ on graphite the binding energy difference between Mo3d$_{5/2}$ and W4f$_{7/2}$ is 229.24 – 32.31 = 196.93 ± 0.04 eV, essentially the same as the core level separation (196.97 ± 0.04 eV) across the HJ stack. This means that "*the supporting graphite substrate is a good common energy reference*" for TMDs when we consider the problem of band alignment. In fact, if we take directly the VBM measured in STS for SL-TMD on graphite (-1.05 eV for WSe$_2$ and -1.84 eV for MoS$_2$), a VBO of 0.79 eV is deduced when we choose graphite as the energy reference. Why does this work? If we treat SL-TMD on graphite as a semiconductor-semimetal heterojunction, then the measured VBM position of SL-TMD on graphite also represents the VBO of the TMD/graphite heterojunction. If we then use the VBOs determined for SL-MoS$_2$/graphite and SL-WSe$_2$/graphite, and apply the transitivity we will deduce a VBO of 0.79 eV for MoS$_2$/WSe$_2$ heterojunction system, essentially the same as the VBO measured using XPS. We suggest that the reason this transitivity holds is related to the weak Van der Waals interactions between TMDs and graphite and between different TMDs.

One fundamental question to address is whether or not individual layers retain their respective electronic structures even after stacking. The definition heterojunction would be meaningful only if this is true. To address this issue, we have carried out theoretical DFT calculations for the electronic structures of the composite system formed by two dissimilar SL-TMD layers. Shown in Fig. 4 is the case for $MoS_2/WSe_2$ calculated using a $MoS_2(\sqrt{13})/WSe_2(\sqrt{12})$ supercell to accommodate the difference in lattice constants. The direct band gap at the original K point in the isolated $MoS_2$ layer remains at K, while the direct band gap at the original K point in the $WSe_2$ layer is folded to the $\Gamma$ point. As can be seen, the electronic structures can be nicely projected into their respective $MoS_2$ and $WSe_2$ layers that are the same as the isolated layers, confirming the validity of treating the $MoS_2/WSe_2$ as a heterojunction. Moreover, a numerical value of $0.94 \pm 0.1$ eV for the valence band offset is obtained in the DFT calculation which agrees very well with the $0.83 \pm 0.07$ eV measured experimentally.

Theoretically, we have also carried out the calculations for heterostructures with other combinations of TMDs. The interlayer interaction between two TMD layers with the same chalcogen species turns out to be significant enough to change the characteristics of the states at the valence band maximum. As an example, the $MoS_2/WS_2$ heterostructure is presented in Supplementary Information, and the energy bands are shown in Fig. S5. We find that the band offset at the K point remains well defined and appears to be independent of the stacking pattern. However, the interlayer coupling moves the VBM position in the $WS_2$ layer from K to $\Gamma$ point, creating an indirect gap about 0.1 - 0.2 eV smaller than the direct gap[21] (see Supplementary Fig. S5). Thus, for optical property which is dominated by the direct transition at the K point, the band offset concept may remain valid[21,22], but one has to take into account the indirect band gap in transport measurements.

**Discussion**

In summary, by using μ-XPS, in conjunction with scanning tunneling spectroscopy we have shown the capability to determine the band alignment in locally formed TMD heterostructures. We determine a type-II alignment in $WSe_2/MoS_2$ with a VBO of 0.83 ± 0.07 eV and a CBO of 0.76 ± 0.12 eV. We further discover that the TMDs and the supporting graphite also form a semiconductor/semimetal heterostructure such that the transitivity hold for different heterostructures formed between SL-TMDs and TMD/graphite. Theoretical investigations show that the electronic band structure of the stacked TMD bilayers containing different chalcogen species resembles a superposition of the energy bands of individual layers, upholding the definition of semiconductor heterojunction. Quantitative agreement of VBO value is found between theoretical calculations and experimental measurements.

# Methods

**Growth and characterizations of 2D monolayers.** The direct growth of $MoS_2$ or $WSe_2$ monolayer crystal flakes on a sapphire substrate by the vapour-phase reaction has been reported in our previous studies[15,16]. In brief, high purity metal trioxides $MO_3$ (M = Mo, W) was placed in a ceramic boat at the heating center of the furnace. A sapphire substrate was placed in the downstream side adjacent to the ceramic boat. Sulfur or Selenium powder was heated by a heating tape and carried by Ar or $Ar/H_2$ to the furnace heating center. The furnace was then gradually heated from room temperature to desired temperature for reaction. After the reaction process, furnace was naturally cooled down to room temperature. These monolayers were characterized atomic force microscopy (Veeco Dimension-Icon system) and a confocal Raman /photoluminescence system (NT-MDT equipped with a 473 nm laser with the spot size of ~ 0.5 μm).

**Scanning tunneling microscopy.** All STM investigations reported here were acquired at 77 K in a homebuilt low temperature in ultra-high-vacuum (UHV) (base pressure $< 6 \times 10^{-11}$ torr). Electrochemically etched W-tips were cleaned *in-situ* with electron beam bombardment. The tunneling bias is applied on sample. Before STM investigations, the samples were cleaned in-situ by heating it up to 250 °C for an extended time (typically longer than 2 hours). The conductance spectra were taken by using a lock-in amplifier with a modulation voltage of 10 mV and at a frequency of 724 Hz.

**Preparation of stacking layers.** To stack the $MoS_2$ ($WSe_2$) monolayer flakes on the undoped Si substrate, a layer of PMMA thin film was coated on the as-grown $MoS_2$ ($WSe_2$) on sapphire as a transfer supporting layer[23]. After dipping in an aqueous NaOH ($H_2SO_4$) solution, the PMMA-supported $MoS_2$ ($WSe_2$) monolayer was detached from sapphire substrates and transferred to the Si(111) substrate (Sheet resistance > 300 Ohm/sq.), followed by the removal of PMMA using acetone. Various 2D heterostructural stacking films such as $WSe_2$ on $MoS_2$ or $WSe_2/MoS_2$ can be obtained by transferring desired 2D flakes onto the other flake.

**Microbeam scanning photoelectron microscopy.** The synchrotron radiation based scanning photoelectron microscopy (SPEM) system is located at beam line 09A1 of National Synchrotron Radiation Research Center, Taiwan. The system is composed of a 16-channel hemisphere electron energy analyzer (model 10-360, PHI), a sample scanning stage, and Fresnel zone plate optics to focus the monochromatic soft X-ray down to 100 nm. By synchronized raster scanning the sample relative to the focused soft X-ray, the excited photoelectrons (PE) were collected and analyzed by the analyzer. The PE intensity of specific core level line can be converted into a two-dimensional surface chemical state distribution image[24]. SPEM can also be operated in µ-XPS mode to acquire high-resolution PE spectra. Combining the imaging and µ-XPS modes makes SPEM a suitable instrument to study the heterostructure of novel 2D materials[25]. The photon energy utilized in this study was 400 eV which was calibrated every 8 hours with the Au 4f core level line of a clean gold foil electrically contacted with the samples to ensure the measurement reproducibility.

**Theoretical calculations.** First-principles calculations were carried out within density functional theory (DFT) using the Vienna Ab initio simulation package (VASP)[26,27]. The interactions between electrons and ions is described by the projector augmented wave (PAW) method[28], and the exchange-correlation potential is described by the Perdew-Burke-Ernzerhof (PBE) generalized gradient approximation[29], with vdW corrections incorporated with vdW-DF functionals[30]. A vacuum thickness of 25 Å is used in order to eliminate the spurious image interaction in the slab calculation, and the energy cut-off of plane waves is 600 eV. The lattice constants of monolayer $MoS_2$ and $WSe_2$ used in the simulation are 3.17 Å and 3.31 Å, respectively. The interlayer spacing is 6.65 Å between $MoS_2$ and $WSe_2$.

**References**

1.  Radisavljevic, B., Radenovic, A., Brivio, J., Giacometti, V. & Kis, A. Single-layer $MoS_2$ transistors. *Nat. Nanotechnol.* **6**, 147-150 (2011).

2.  Splendiani, A. *et al.* Emerging photoluminescence in monolayer $MoS_2$. *Nano Lett.* **10**, 1271-1275 (2010).

3.  Mak, K. F., Lee, C., Hone, J., Shan, J. & Heinz, T. F. Atomically thin $MoS_2$: a new direct gap semiconductor. *Phys. Rev. Lett.* **105,** 136805 (2010).136805

4.  Wu, S. F. *et al.* Electrical tuning of valley magnetic moment through symmetry control in bilayer $MoS_2$. *Nat. Phys.* **9**, 149-153 (2013).

5.  Xiao, D., Liu, G. B., Feng, W. X., Xu, X. D. & Yao, W. Coupled spin and valley physics in monolayers of $MoS_2$ and other Group-VI dichalcogenides. *Phys. Rev. Lett.* **108**, 196802 (2012)

6.  Britnell, L. *et al.* Strong light-matter interactions in heterostructures of atomically thin films. *Science* **340**, 1311-1314(2013).

7.  Hong, X. P. *et al.* Ultrafast charge transfer in atomically thin $MoS_2/WS_2$ heterostructures. *Nat. Nanotechnol.* **9**, 682-686, doi:Doi 10.1038/Nnano.2014.167 (2014).

8.  Withers, F. *et al.* Light-emitting diodes by band-structure engineering in van der Waals heterostructures. *Nat. Mater.* **14**, 301-306, doi:10.1038/nmat4205 (2015).

9.  Fang, H. et al. Strong interlayer coupling in van der Waals heterostructures built from single-layer chalcogenides. *Proc. Natl. Acad. Sci. USA* **111**, 6198-6202(2014).

10. Franciosi, A. & Van de Walle, C. G. Heterojunction band offset engineering. *Surf. Sci. Rep.* **25**, 1-140(1996).

11. Bratina, G., Sorba, L., Antonini, A., Biasiol, G. & Franciosi, A. AlAs-GaAs heterojunction engineering by means of Group-IV elemental interface layers. *Phys. Rev. B* **45**, 4528-4531(1992).

12. Kowalczyk, S. P., Cheung, J. T., Kraut, E. A. & Grant, R. W. CdTe-HgTe ($\bar{1}\bar{1}\bar{1}$) heterojunction valence-band discontinuity: a common-anion-rule contradiction. *Phys. Rev. Lett.* **56**, 1605-1608 (1986).

13. Shih, C. K. & Spicer, W. E. Determination of a natural valence-band offset: the case of HgTe- CdTe. *Phys. Rev. Lett.* **58**, 2594-2597 (1987).

14. Cong, C. X. *et al.* Synthesis and optical properties of large-area single crystalline 2D semiconductor $WS_2$ monolayer from chemical vapor deposition. *Adv. Opt. Mater.* **2**, 131-136 (2014).

15. Huang, J. K. *et al.* Large-area synthesis of highly crystalline $WSe_2$ mono layers and device applications. *ACS Nano* **8**, 923-930 (2014).

16. Lee, Y. H. *et al.* Synthesis of large-area $MoS_2$ atomic layers with chemical vapor deposition. *Adv. Mater.* **24**, 2320-2325 (2012).

17. Jin, W. C. *et al.* Direct measurement of the thickness-dependent electronic band structure of

MoS$_2$ using angle-resolved photoemission spectroscopy. *Phys. Rev. Lett.* **111**, 106801 (2013).

18. Tonndorf, P. *et al.* Photoluminescence emission and Raman response of monolayer MoS$_2$, MoSe$_2$ and WSe$_2$. *Opt. Express* **21**, 4908-4916 (2013).

19. Zhang, C. D., Johnson, A., Hsu, C. L., Li, L. J. & Shih, C. K. Direct imaging of the band profile in single layer MoS$_2$ on graphite: quasiparticle energy gap, metallic edge states and edge band bending. *Nano Lett.* **14**, 2443–2447(2014).

20. Zhang, C.D. *et al.* Measuring critical point energies in transition metal dichalcogenides. arXiv:1412.8487 (2014).

21. Pasqual, R. *et al.* Observation of long-lived interlayer excitons in monolayer MoSe$_2$-WSe$_2$ heterostructures. *Nat. Commun.* doi: 10.1038/ncomms7242 (2015)

22. van der Zande, A. M. *et al.* Tailoring the electronic structure in bilayer molybdenum disulfide via interlayer twist. *Nano Lett.* **14**, 3869-3875(2014).

23. Lebegue, S. & Eriksson, O. Electronic structure of two-dimensional crystals from ab initio theory. *Phys. Rev. B* **79**, 115409 (2009).

24. Chiou, J. W. X-rays in nanoscience: spectroscopy, spectromicroscopy, and scattering techniques Ch. 4. (Weinheim : Wiley-VCH ; Chichester, 2010).

25. Shiu, H. W. *et al.* Graphene as tunable transparent electrode material on GaN: Layer-number-dependent optical and electrical properties. *Appl. Phys. Lett.* **103**, 081604 (2013).

26. Kresse, G. & Furthmuller, J. Efficiency of ab-initio total energy calculations for metals and semiconductors using a plane-wave basis set. *Comput. Mater. Sci.* **6**, 15-50 (1996).

27. Kresse, G. & Furthmuller, J. Efficient iterative schemes for ab initio total-energy calculations using a plane-wave basis set. *Phys. Rev. B* **54**, 11169-11186 (1996).

28. Blochl, P. E. Projector augmented-wave method. *Phys. Rev. B* **50**, 17953-17979 (1994).

29. Perdew, J. P., Burke, K. & Ernzerhof, M. Generalized gradient approximation made simple. *Phys. Rev. Lett.* **77**, 3865-3868 (1996).

30. Klimes, J., Bowler, D. R. & Michaelides, A. Van der Waals density functionals applied to solids. *Phys. Rev. B* **83**, 195131 (2011).

**Acknowledgements**

This research was supported by Academia Sinica, National Science Council Taiwan (NSC-102-2119-M-001-005) and AFOSR BRI. The work at UT Austin is support with grants from the Welch Foundation (F-1672), and the US National Science Foundation (DMR-1306878), and the work at Georgia Tech is supported by the US Department of Energy, Office of Science, Basic Energy Sciences, under Award No. DE-FG02-97ER45632. C.K.S also thanks the National Science Council, Taiwan for financial supports for a visiting chair professorship at the National Tsing Hua University, Taiwan (NSC 102-2811-M-007-034).

**Author Contributions**

C.K.S. and L.J.L. conceived the experiment. L.J.L. coordinated the growth and XPS effort, C.K.S. coordinated the STM effort, and M.Y.C. directed the theoretical simulation effort. C.H.C. and C.Y.C. performed the growth and M.H.C. prepared the CVD stacked films and carried out Raman and PL characterizations. H.W.S. and C.H.C. performed the XPS measurement. C.D.Z. carried out the STM measurement. C.D.Z. and C.K.S. analysed the STM data. C.P.C. carried out the theoretical simulation. L.J.L., C.K.S. and M.Y.C. wrote the papers with inputs from the other co-authors.

**Additional information**

Reprints and permissions information is available online at www.nature.com/reprints.

**Competing financial interests**

The authors declare no competing financial interests.

**Figure Legends**

**Figure 1 | Optical microscopy images and photoluminescence spectroscopy taken on the staked WSe$_2$/MoS$_2$ heterostructure. a,** Optical micrograph (OM) and atomic force microscopy (AFM) images for the WSe$_2$/MoS$_2$ heterostructural stacked flakes on a sapphire substrate. **b**, Photoluminescence spectra for the selected sites including MoS$_2$ only (A), WSe$_2$ only (C) and WSe$_2$/MoS$_2$ (B) stacked areas.

**Figure 2 | μ-XPS measurements on the stacked MoS$_2$/WSe$_2$ and WSe$_2$/MoS$_2$ heterostructures. a,** The Mo3d and W4f mappings for the same physical area. The right figure is the overlapped mapping which allows the identification of MoS$_2$/WSe$_2$ stacked areas. A and B points are the typical stacking areas where the XPS are taken. **b,** The spectra for the selected insulated WSe$_2$ and MoS$_2$ flakes, the stacked MoS$_2$/WSe$_2$ and WSe$_2$/MoS$_2$ heterostructure. All numbers are quoted in electron volt. There is an uncertainty of ± 0.04 eV when determining the energy levels for all peaks.

**Figure 3 | STM images and the tunneling spectra of WSe$_2$ and MoS$_2$ grown on HOPG. a,** STM image zoomed in on the edge of WSe$_2$ grown on HOPG. The inset shows the atomic resolution image taken on the SL-WSe$_2$. For the inset, U = - 1.4 V, I = 10 pA. **b,** The *dI/dV-V* spectrum taken on the SL-WSe$_2$ flake. It is plotted in both the linear scale (upper panel) and the logarithmic scale (lower panel). The green dashed arrows indicate the positions of the local VBM at Γ and K points, which are equal to -1.65 eV and -1.05 eV, respectively. The CBM is assigned at +1.03 eV. **c,** The clear threshold, corresponding to VBM at K, can be seen in the *dI/dV* spectrum taken with much more close tip-sample distance (~3.5 Å closer than **b**). For comparison, the spectra taken with the stabilization bias with the gap (-0.8 V) is shown in black, while the one taken on bare graphite is in blue. The similar *dI/dV (in black) and "close-in" (red in inset)* spectra of SL-MoS$_2$ are displayed in **d**, while the valence band maxima at Γ and K are marked in the inset. The *E$_C$* (blue arrow) corresponds to the CBM. **e,** The diagram of band alignments between single layer MoS$_2$ and WSe$_2$. The local VBM at Γ and global VBM at K are shown in cyan and red, respectively.

**Figure 4 | First-principles calculations for the electronic structures of the composite system formed by two dissimilar SL-TMD layers. a,b,** Energy band structure of the MoS$_2$/WSe$_2$ bilayer calculated using a supercell containing rotated $\sqrt{13} \times \sqrt{13}$ and $\sqrt{12} \times \sqrt{12}$ unit cells of MoS$_2$ and WSe$_2$, respectively, in order to minimize the strain in individual layers due to lattice mismatch. The projected bands onto Mo and W atoms are shown in **a** and **b**, respectively with the amount of Mo (W) projection represented by the size of blue (red) circles. **c,** One-dimensional charge density (integrated over the horizontal direction) illustrating that the states at the band edges belong to a distinct layer.

Figure 1

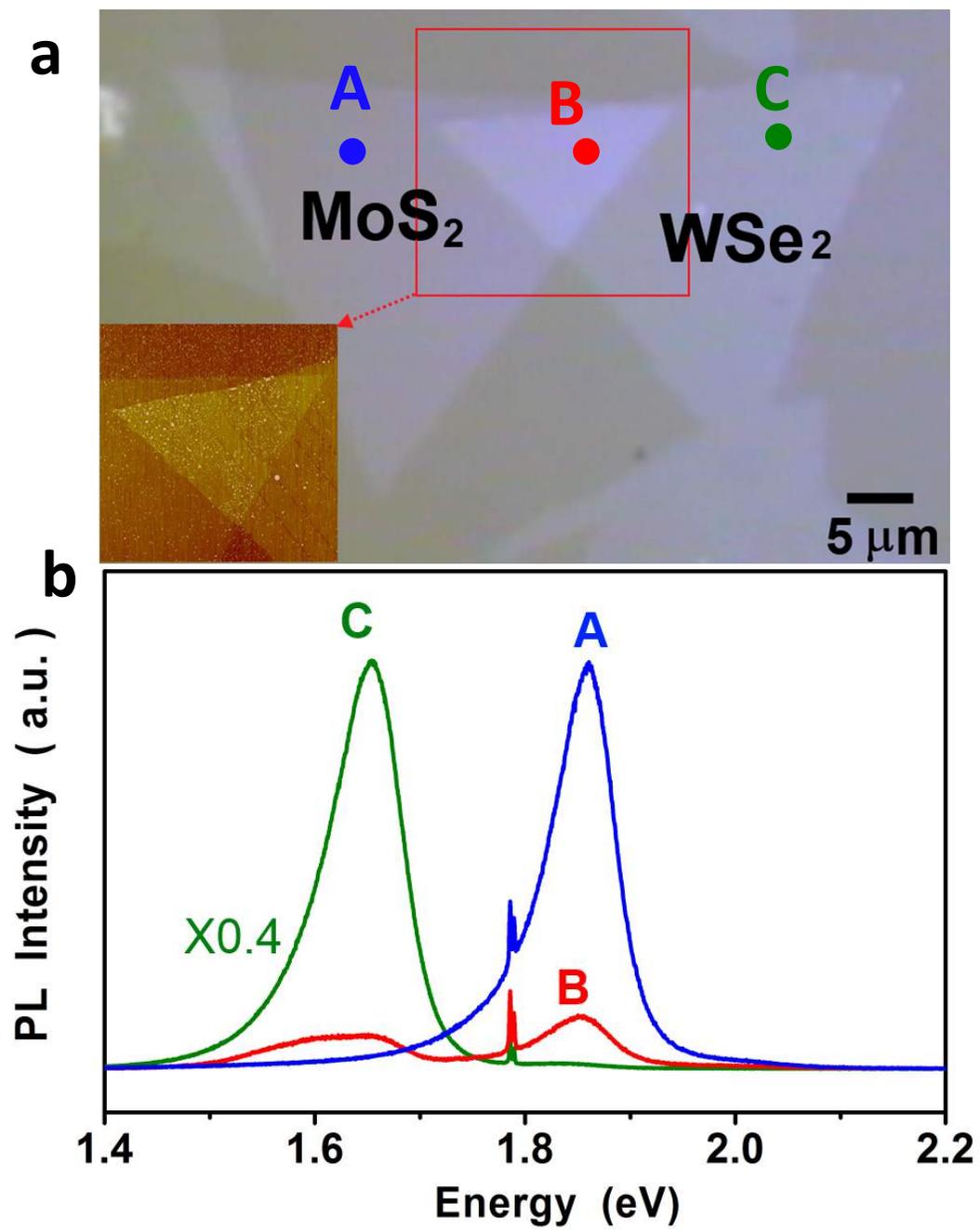

Figure 2

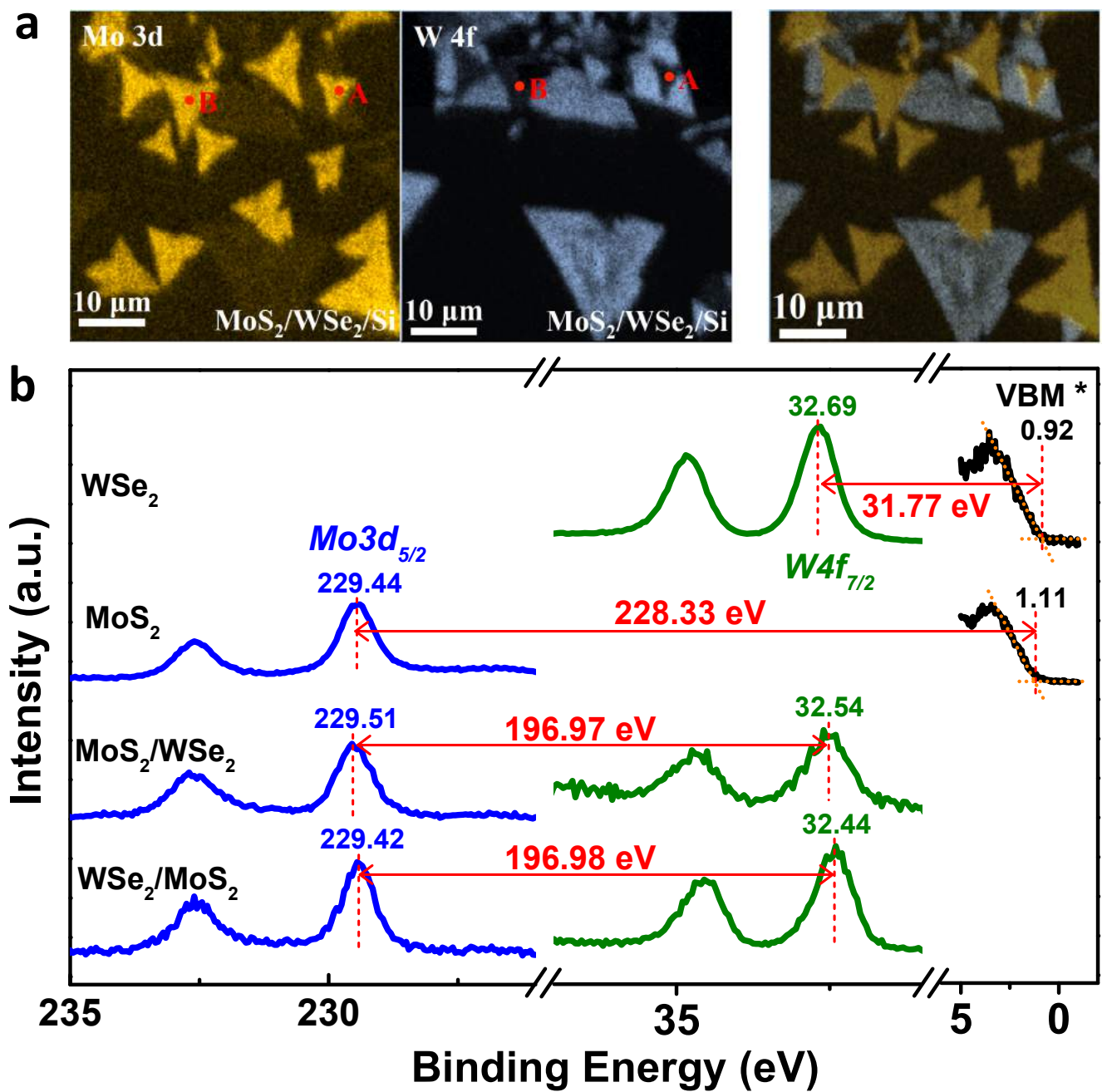

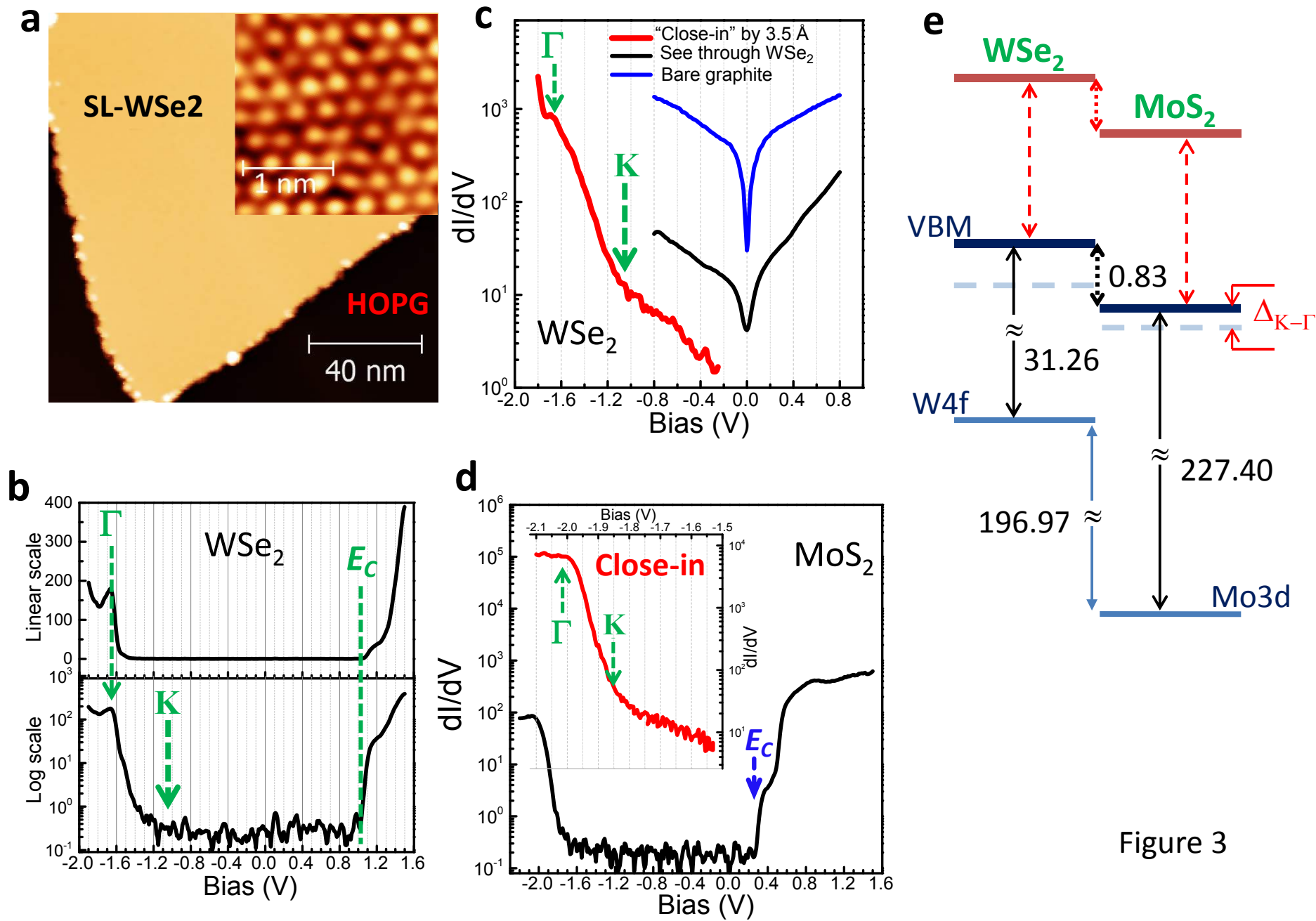

Figure 3

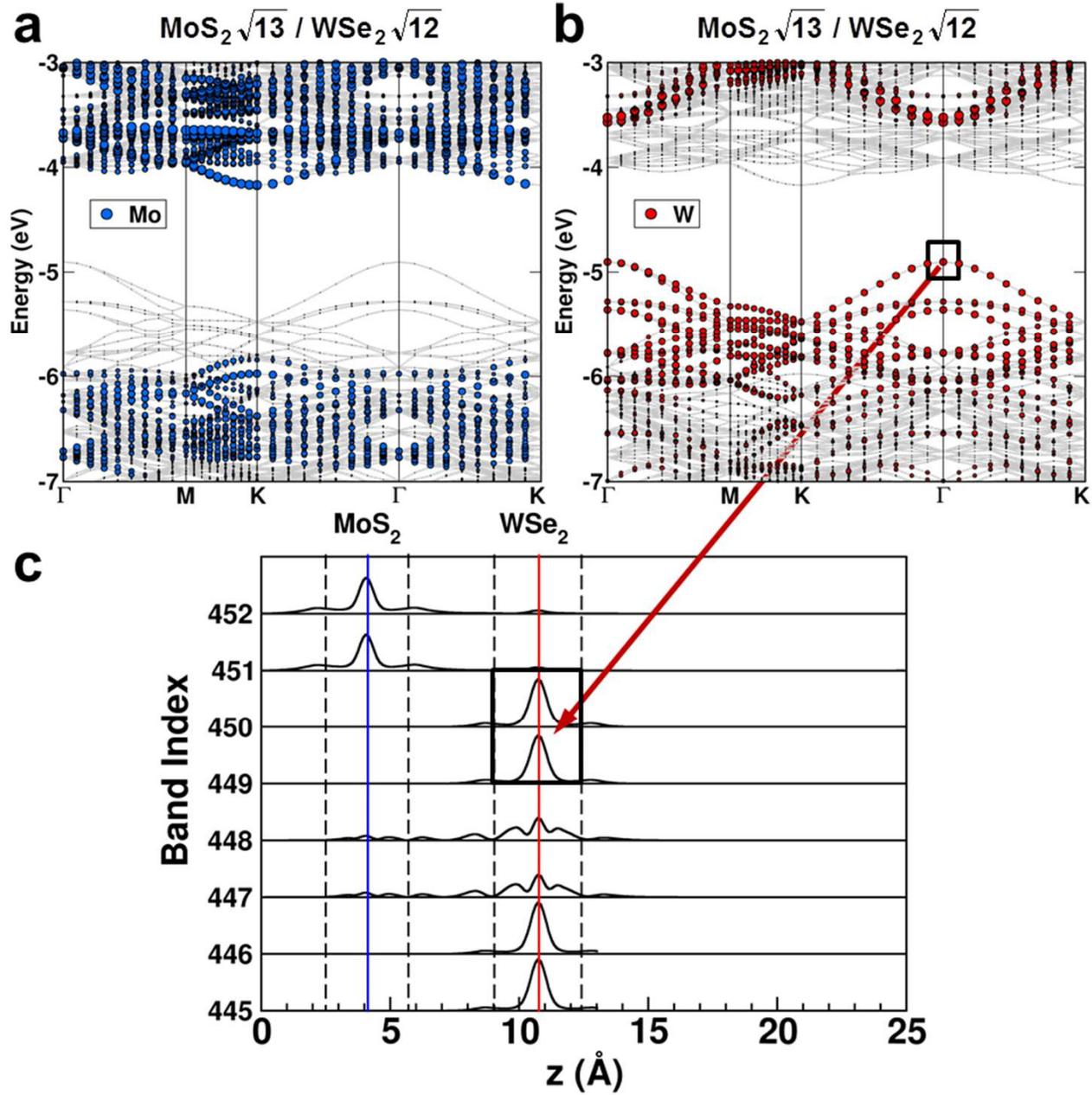

Figure 4

# Supplementary Information: Determination of band alignment in the single layer MoS$_2$/WSe$_2$ heterojunction

*Ming-Hui Chiu[1,^], Chendong Zhang[2,^], Hung Wei Shiu[3], Chih-Piao Chuu[1], Chang-Hsiao Chen[1], Chih-Yuan S. Chang[1], Chia-Hao Chen[3], Mei-Yin Chou[1,4,5], Chih-Kang Shih[2,*] and Lain-Jong Li[1,6,*]*

[1]*Institute of Atomic and Molecular Sciences, Academia Sinica, No. 1, Roosevelt Rd., Sec. 4, Taipei, 10617, Taiwan*
[2] *Department of Physics, University of Texas at Austin, Austin, TX 78712, USA*
[3] *National Synchrotron Radiation Research Center, HsinChu 30076, Taiwan*
[4] *School of Physics, Georgia Institute of Technology, Atlanta, GA 30332, USA*
[5]*Deapartment of Physics, National Taiwan University, Taipei 10617, Taiwan*
[6]*Physical Sciences and Engineering Division, King Abdullah University of Science and Technology, Thuwal, 23955-6900, Kingdom of Saudi Arabia.*

*^These authors contributed equally.*
*\*Corresponding author E-mail: lanceli@gate.sinica.edu.tw, shih@physics.utexas.edu*

**Figure S1 Characterizations of synthetic transition metal dichalcogenides monolayers on sapphire substrates.**

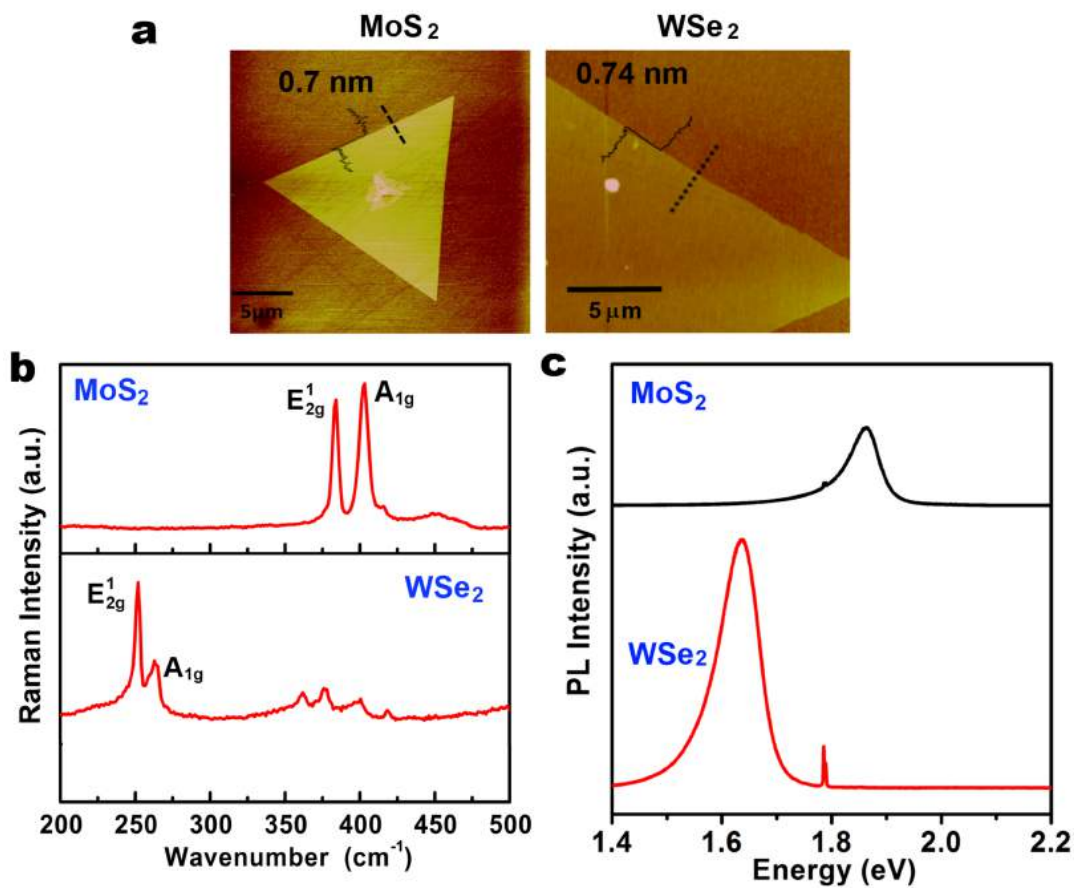

**a,** The thickness obtained from AFM cross-sectional profiles is around 0.7 to 0.8 nm, indicating that these as-grown flakes are single-layered. **b.** Raman and **c.** PL spectra for the synthetic $MoS_2$ and $WSe_2$ flakes. These spectroscopic features are consistent with those obtained from exfoliated single layers as shown in Table S1. The measurements were performed in a confocal Raman/photoluminescence system equipped with a 473 nm laser with the spot size of ~ 0.5 μm.

# Determination of VBM* using XPS

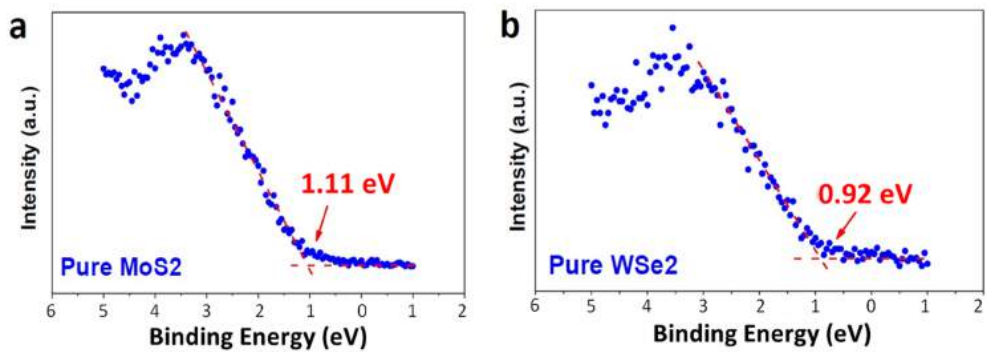

**Figure S2**

Figure S2 illustrates how the VBM location is determined using XPS. Here the leading edge is extrapolated to intersect with the background base line. This procedure has been used to deduce the VBM position of conventional semiconductors whose VB DOS have very similar shape. The result is consistent with a more sophisticated least square fit with respect to broadened DOS. This procedure, however encounters some difficulties in determine the actual VBM position in TMD materials. In SL-TMDs, the global VBM is located at the K point.

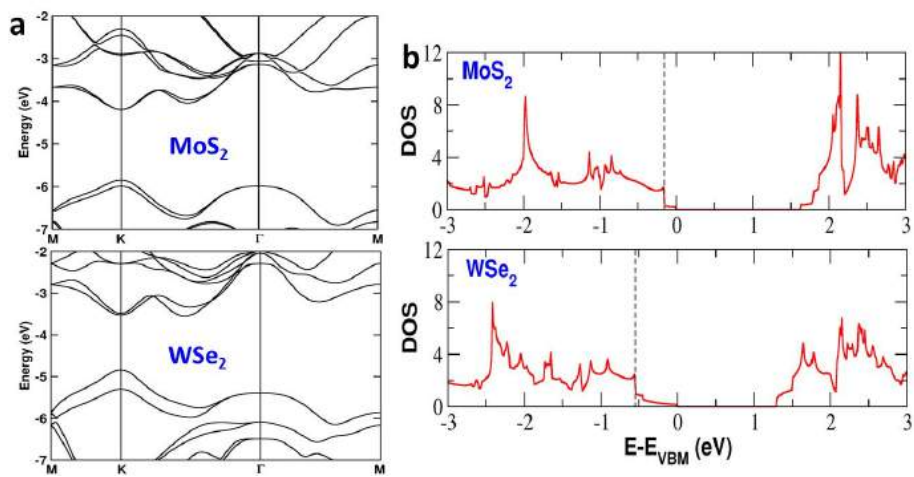

**Figure S3**

However near the VBM, the DOS is dominated by the states near the $\Gamma$ point (at the zone center)

as illustrated in Fig. S3b which contains the theoretical calculation of DOS of two TMDs ($MoS_2$ and $WSe_2$) using DFT. The three dashed vertical lines mark the energy positions of Γ points in three SL-TMD compounds respectively. The much larger contribution of DOS near the Γ point in comparison to that near the K point is due to a larger effective mass at the Γ point (by a factor of 4-5) and the double degeneracy. Since XPS cannot resolve the Γ-K splitting, the fitting of leading edge will consistently yield a VBM position very close to the local VBM at the Γ point, but offset by ~ 0.2 eV above it. Since VBO is referred to as the difference between the VBM positions, this offset will be cancelled out when VBO is deduced. Thus, the VBO deduced from the XPS measurement will correspond to the band offset of the local VBM at the Γ point (referred to as VBO*). As discussed in the main text, STS allows us to resolve Γ-K splittings which can be added to the VBO* to obtain accurate values of VBOs.

**Figure S4 μ-XPS measurements of SL MoSe$_2$ and WSe$_2$ on HOPG**

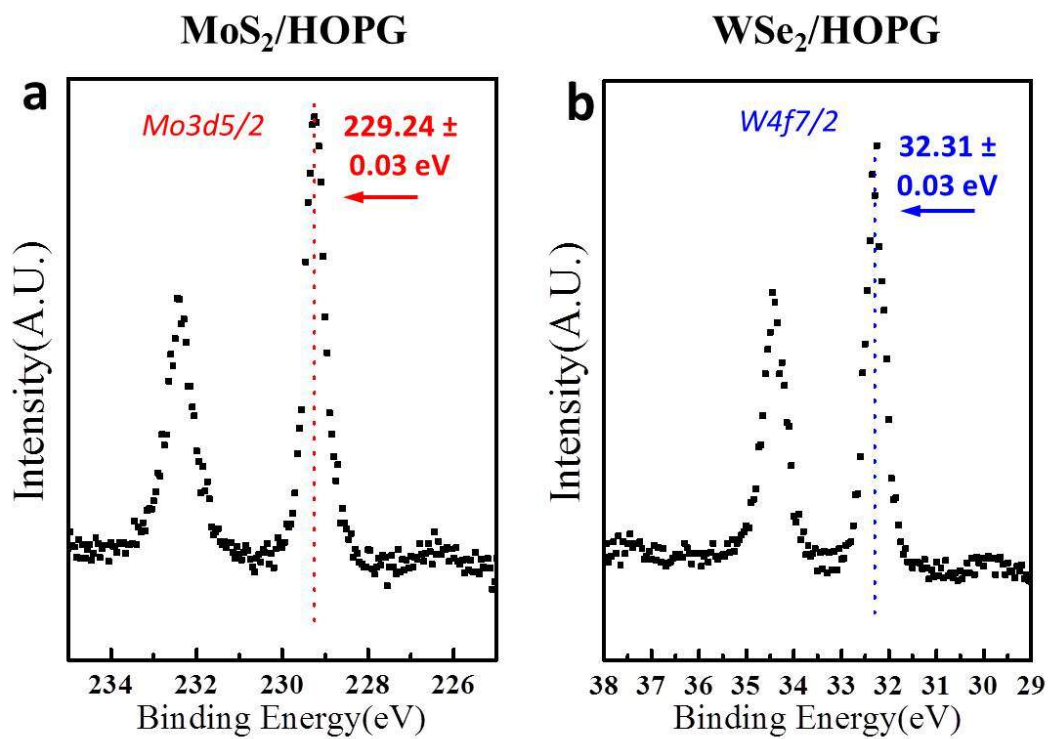

The μ-XPS taken on the SL MoS$_2$ and WSe$_2$ grown on HOPG. The measurements yield a binding energy of 229.24 ± 0.03 eV for Mo3d$_{5/2}$ (shown in **a**), corresponding to a separation of 228.40 eV to the true VBM (-1.84 eV measured by STS) in MoS$_2$. Similarly W4f$_{7/2}$ in SL-WSe$_2$ on graphite has a binding energy of 32.31 ± 0.03 eV (shown in **b**), corresponding to a value of 31.26 eV when referenced to the true VBM (-1.05 eV measured by STS).

**Theoretical calculations for $MoS_2/WS_2$**

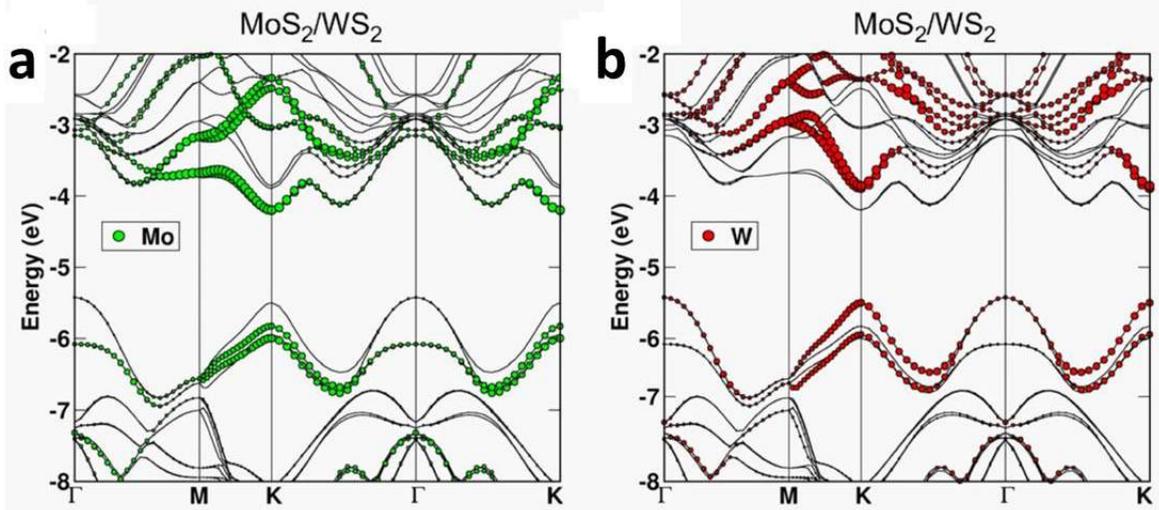

**Figure S5**

The electronic structure of the $MoS_2/WS_2$ bilayer exhibits additional features as shown in Fig. S5a and b where the amount of Mo (W) projection is represented by the size of green (red) circles. The two layers have same chalcogen atoms and almost identical lattice constants. We have performed the calculations for a few stacking patterns and found that the band offset at the K point remains well defined and appears to be independent of the stacking pattern. However, the interlayer coupling moves the VBM position in the $WS_2$ layer from K to Γ point, creating an indirect gap about 0.1 - 0.2 eV smaller than the direct gap [ref 1]. Note that our calculation of $WS_2$ on $MoS_2$ corresponds to a 60-degree (2H) stacking, lattice matched bilayer (namely an anti-aligned stacking). We note that a recent optical study of the $MoS_2$ bilayer system with artificial designed stacking angles shows that the interlayer coupling depends strongly upon the twist angle, with the coupling strength significantly reduced at twist angles between 0 and 60 degree [ref 3]. Since in our system of $WS_2$ on $MoS_2$, the VBM at the Γ point is about 0.1 - 0.2 eV above the K-point in contrast to a value of 0.4 - 0.5 eV for the bilayer, it remains to be seen whether the weakened interlayer coupling in the incoherent stacking will recover the direct gap characteristics.

**Reference:**

1.  Pasqual, R. et al. Observation of long-lived interlayer excitons in monolayer $MoSe_2$-$WSe_2$ Heterostructures. *Nat. Commun.*, doi: 10.1038/ncomms7242.

2.  van der Zande, A. M. et al. Tailoring the electronic structure in bilayer molybdenum disulfide via interlayer twist. *Nano Lett.* **14**, 3869-3875 (2014).

2.  Lebegue, S. & Eriksson, O. Electronic structure of two-dimensional crystals from *ab initio* theory. *Phys. Rev. B* **79**, 115409 (2009).

**Table S1 Summary of Raman and PL measurements for exfoliated and CVD grown SL-TMDs.**

|  | Exfoliated | | | CVD grown | | |
|---|---|---|---|---|---|---|
|  | Raman (1/cm) | | PL (eV) | Raman (1/cm) | | PL (eV) |
|  | $E_{2g}$ | $A_{1g}$ |  | $E_{2g}$ | $A_{1g}$ |  |
| $MoS_2$ | 384 | 403 | 1.9 | 384 | 403 | 1.85 |
| $WSe_2$ | 249 | 261 | 1.65 | 248 | 259 | 1.63 |